\documentclass[12pt]{iopart}

\usepackage{iopams}  
\usepackage{graphicx}
\usepackage{tikz}
\usepackage{tikz-feynman}
\usepackage{feynmp}
\usepackage{subfigure}
\usepackage{wrapfig}
\begin{document}

\title[EFT Calculation of LIGO-like Compton Scattering]{Effective Field Theory Calculation of LIGO-like Compton Scattering}

\author{Noah M. MacKay
\\ ORCID ID: 0000-0001-6625-2321}

\address{Institut für Physik und Astronomie, Universit\"at Potsdam,\\ Karl-Liebknecht-Str. 24/25, 14476 Potsdam, Germany}
\ead{noah.mackay@uni-potsdam.de}
\vspace{10pt}
\begin{indented}
\item[]\today
\end{indented}

\begin{abstract}
We use effective field theory (EFT) to calculate the scattering amplitude of a LIGO-like graviton-scalar Compton interaction. We gauge the center-of-momentum energy $\sqrt{s}$ between one gravitational-wave (GW) graviton (one quantum of the coherent bulk of an astrophysical GW, with energy $E_g=\hbar\omega_\mathrm{GW}$) and a resting heavy target (a suspended mass with rest energy $E_M=m_Mc^2$ found in laser interferometer observatories) to be of order $\sim10^{1.5}$ PeV -- at the energy scale within the extremes of astroparticle physical phenomena. This back-of-the-envelope calculation supports the calculation of a convergent cross section in our LIGO-like Compton analysis, which we indeed recover using standard EFT Feynman rules and relevant traceless-transverse gauges for the graviton polarizations. We obtain that the cross section $\sigma$ is largely dependent on the center-of-momentum energy, and from this we define the corresponding impact parameter via $\sigma=\pi b^2$. This impact parameter, after coherence-state population enhancement, scales with GW energetics along with the unique coupling (in natural units) $\sqrt{\omega_\mathrm{GW}/m_M}\sim10^{-21}$ -- the same order of magnitude as astrophysical GW strain. One finds furthermore that, after isolating the GW energetics sector from the GW-mirror impact parameter, the revised length scale $\tilde{b}/(GM)\approx1.76\pi$ quantifies the pre-merger stage of compact binary coalescence, which is compared with $b/(GM)>14$ calculated from the early-inspiral worldline quantum field theory framework. Conventional insight of GW-mirror response is recovered, such that the impact parameter $b$ scales exactly with the mirror recoil $\Delta L\sim10^{-18}\,\mathrm{m}$ after having made contact with the GW.
\end{abstract}

%
\noindent{\it Keywords}: Gravitational waves, Effective field theory, Compton scattering\\
%
\submitto{\CQG}
%
\maketitle
%
%

\section{Introduction} \label{intro}

Since their discovery on September 14, 2015, gravitational waves (GWs) have been routinely observed by the LIGO-VIRGO-KAGRA (LVK) collaboration \cite{GWOSC, LIGOScientific:2018mvr, LIGOScientific:2021usb, KAGRA:2021vkt, LIGOScientific:2025slb}, now called the International Gravitational Wave Network (IGWN). Astrophysical sources of GWs have been well understood to be coalescing compact binaries (CCBs), which send out GW signals encoding the quintessential chirp rise (the dynamic frequency and amplitude enhancement from inspiral up to merger) and the post-merger ringdown phase with possible after-merger rotations and an exponential damping towards a zero flat-line. While these waveforms are implementented computationally via numerical relativity (e.g. \texttt{SEOBNRv4} \cite{Bohe:2016gbl} and \texttt{v5} \cite{Khalil:2023kep}, \texttt{IMRPhenom} \cite{Ajith:2007qp}, and the \texttt{SEOBNRv4T} tidal deformation extension \cite{Hinderer:2016eia, Steinhoff:2016rfi, Steinhoff:2021dsn} among many more waveform families), semi-analytically via the effective one-body (EOB) model \cite{Buonanno:1998gg, Buonanno:2005xu}, and approximately via the more recent mass-shell model \cite{MacKay:2024qxj, MacKay:2025uyg}, their sources are behaviorally quadrupolar. 

Via the linearized Einstein field equations (EFEs) in a weak field limit, and modeling the CCB as two point-mass constituents, one obtains e.g. the circular traceless-transverse (TT) gauged spatial waveforms (using $c=1$): 
\begin{equation}\label{gwsourced}
 h^\mathrm{TT}_{ij}=-\frac{4G\mu L^2\Omega^2}{D}\exp(i\,2\Omega t)\varepsilon^\mathrm{TT}_{ij},
\end{equation} 
where $D$ is the luminosity distance between the observer and the source, $\mu=m_1m_2/M$ is the reduced mass of the binary ($M=m_1+m_2$ is the total mass), $L$ is the binary separation, $\Omega$ is the orbital frequency,  $h_+\propto\cos(2\Omega t)$ and $h_\times\propto\sin(2\Omega t)$ via the $3\times3$ TT-gauged matrix $\varepsilon^\mathrm{TT}_{ij}$, and the strain amplitude is distinctly defined in source-dependent values. One must keep in mind, however, that realistic CCBs complicate the simple waveforms by incorporating dynamic post-Newtonian (PN) \cite{Blanchet:2013haa} and perturbative post-Minkowskian (PM) \cite{Damour:2016gwp} corrections, leading to the inspiral-merger-ringdown (IMR) behavior described earlier. 

In contrast, GWs propagating in the vacuum act as free waves, analogously to free electromagnetic waves in the vacuum. Derived from the linearized EFEs with no quadrupolar source (i.e., taking the identical steps to derive Eq. (\ref{gwsourced}) however for $T_{\mu\nu}=0$), TT-gauged GWs in the vacuum follow the d'Alembert wave equation and yield plane wave solutions:
 \begin{equation}\label{gwvacuum}
h^\mathrm{TT}_{\mu\nu}=A_0\exp(ik_\alpha x^\alpha)\varepsilon^\mathrm{TT}_{\mu\nu}.
\end{equation}
Comparing Eqs. (\ref{gwsourced}) and (\ref{gwvacuum}), $A_0$ denotes the wave amplitude relating to the strain and source-dependent factors, $k_\mu=(\omega,-\vec{k})$ is the 4-vector for the wave number which is in the inner product with $x^\mu=(t,\vec{x})$, using the metric signature $\eta_{\mu\nu}=\mathrm{diag}(+,-,-,-)$ (see footnote\footnote{This metric signature is rather unconventional in general relativity, but it is widely used in particle physics and Q/EFT.}), and $\varepsilon^\mathrm{TT}_{\mu\nu}$ is the $4\times4$ TT-gauged polarization tensor. Therefore, one observes that a correspondence arises between Eqs. (\ref{gwsourced}) and (\ref{gwvacuum}) -- between GWs from a source and GWs propagating in vacuum. Extending this correspondence to connect the two as one in the same, we may define any GW propagating in the vacuum as a GW originated from a quadrupolar source. This is essential for the few hundred GW events detected across the IGWN, especially when uncovering source parameter information from the obtained data. 

This GW correspondence is also essential in the context of particle physics and effective field theory (EFT). Just as free electromagnetic waves are quantized as propagating photons in an incoherent superposition, GWs in a vacuum (albeit originating from a source) are ideally quantized as propagating coherent-state gravitons despite their hypothetical nature (see e.g. Refs. \cite{Feynman:2002, Goldberger:2004jt, Goldberger:2006bd, Goldberger:2007hy, Kol:2007bc, Goldberger:2009qd, Foffa:2013qca, Rothstein:2014sra, Porto:2016pyg, Levi:2018nxp, Rafie-Zinedine:2018izq, Mogull:2020sak, Aoki:2024boe}). Similar to photons, gravitons are massless (see Ref. \cite{Nakanishi:1979fg}) and chargeless; unlike spin-1 photons, gravitons are spin-2 quanta, following the tensoral rank of the metric perturbation $h_{\mu\nu}$, and are self-interacting (a quality shared with spin-1 gluons). It is these self-interactions that introduce nonlinear, higher-ordered corrections that directly follow conventional PM expansion.

Exact quantization of GWs -- and the gravitational interaction -- as gravitons has been a long-standing question in physics that served as the engine behind many quantum gravity and quantum-classical-correspondence (QCC) theories. Issues arise, however, from infrared divergences and therefrom the tentative inability to reconcile general relativity (GR) with quantum mechanics. Because they remain (as of \today) hypothetical, gravitons are often used as analytical tools in current gravitational EFTs, and in particular for EFT modeling of binary coalescence and its associated GW radiation. The worldline quantum field theory (WQFT) formalism \cite{Mogull:2020sak}, for instance,  and its subsequent developments e.g. \cite{Driesse:2024feo}, attempt to describe the PN regime of coalescence using Feynman diagrams, calculating scattering amplitudes from which end-state measurables e.g. relative scattering angle and inspiral-phase radiated energy may be derived \cite{Driesse:2024feo}. 

Within this logic, the merger phase in IMR dynamics can be effectively described as either a $2\rightarrow1$ (i.e. $\mathrm{CCB}\rightarrow\mathrm{BH}$) inelastic collision channel or a $2\rightarrow2$ (i.e. $\mathrm{CCB}\rightarrow\mathrm{BH+radiation}$) elastic collision channel (see e.g. Ref. \cite{Aoki:2024boe}). EFT approaches to the late-inspiral/pre-merger stages of coalescence beyond Ref. \cite{Aoki:2024boe} are however a particularly cumbersome task, as one must take a strongly-curved background into account. Therefore, to properly approach on-shell IMR dynamics, one must draft an EFT that is similar in methodology to a QFT on curved spacetimes (see e.g. \cite{Birrell:1982ix, Bunch:1979uk, Bekenstein:1981xe, Alsing:2000ji} for related literature on curved-spacetime-QFTs). While this is a topic of particular interest, this lies beyond the scope of this work.

Rather surprisingly, there are seldom EFT approaches to LIGO-like detection events, i.e. the interaction between an incoming GW and a suspended mass in laser interferometer observatories. Due to the success the Standard Model and QFTs have in predicting end-state observables from a particle-physics experiment done in e.g. CERN, one may argue that an EFT perspective on an LVK detection -- or more relevantly an eventual detection made by the Einstein Telescope (ET) \cite{ET:2025xjr} or the Laser Interferometer Space Antenna (LISA) \cite{LISA:2017pwj} -- is just as useful and informative. It is in this spirit that we proceed here. 

\subsection{Using EFT to depict GW detection}

We note, of course, the differences between a high-energy scattering experiment and a LIGO-like detection. We remind that, in comparison to a relativistic heavy-ion collision taken place in e.g. CERN, a laser interferometer detection involves a coherent GW interacting with a localized probe. Given the hypothetical understanding of GWs as a coherence state of gravitons (presumptively having a large occupation number $N$), we consider the essential Compton scattering a single graviton within the coherent bulk undergoes with a LIGO-like detector apparatus. To do so, we revisit quintessential graviton-scalar Compton scattering at tree level for a simple analysis. Here, the scalar particle models the hanging mass, which in the center-of-momentum frame appears to be in motion with a small recoil while remaining deeply non-relativistic.

We further acknowledge that gravitational Compton scatterings have been previously approached, albeit primarily focusing on the low-energy (infrared) structure of the resulting scattering amplitude and cross section \cite{Holstein:2017dwn}, see footnote\footnote{Ref. \cite{Holstein:2017dwn} found that the low-energy regime of Compton scattering between a single graviton and a scalar particle yields a divergent cross section.}. We clarify that our approach to the process is not intended to represent a literal scattering event between a graviton in the detector. We show that the relevant scattering amplitude and total cross section is inconceivably small for a single graviton. Rather, it serves as a matching calculation encoding the graviton-matter coupling that underlies the EFT description, which we may enhance (whenever relevant) with the coherent-state occupation number $N$ to effectively describe the GW-mirror interaction. In this work, the relevant kinematic invariant (the center-of-momentum energy) of the graviton-matter interaction is of order $\sim31.6$ PeV (to give a glimpse of Eq. [\ref{cmeng}]), which reflects an auxiliary energy scale -- that of the extremes of astroparticle physical phenomena -- of the underlying amplitude we aim to calculate. However, the hierarchy where the graviton quantum energy is far less than the rest energy of the suspended mass, $E_\mathrm{g}\ll m_M$, places the system in a soft-graviton, heavy-target EFT regime -- closely replicating the physical conditions of a typical LIGO-like experiment.

From this analysis, it is our goal to use EFT methods to approach LIGO-like events, deriving the detector response from the underlying graviton-matter interactions. In particular, we aim to obtain the total scattering amplitude of our Compton scattering interaction (in both center-of-momentum and laboratory frames), and show how classical observables emerge in the coherent (classical) limit. From this EFT-derived scattering amplitude, we also aim to obtain an expression for the total cross section $\sigma=\pi b^2$ and the corresponding impact parameter $b$, along with its scale $b/(GM)$ if appropriate. The latter calculation for $b/(GM)$ is to compare with e.g. WQFT that can calculate $b/(GM)>14$ \cite{Driesse:2024feo} for early-inspiral two-body gravitational scattering -- the calculation in our context is to recover near-merger intuition if one isolates the GW energetics sector of the GW-mirror parameter.

The paper is structured in the following layout: in Section \ref{sec2} we provide a review of the essentials behind QFT and EFT for comprehension, including the construction of the relevant Lagrangian and necessary Feynman rules for this Compton scattering analysis in Section \ref{feynrules}. In Section \ref{scatt}, we calculate basic tree-level Compton scattering diagrams in the center-of-momentum frame, to solve for their respective and the total amplitudes. With these results we further compute the magnitude of the total amplitude in the laboratory frame in Section \ref{tamp}: the frame in which remote observers detect any signal inferred from LVK observation runs. It is with this amplitude that we solve for the total cross section and the impact parameter. After a discussion of our results in Section \ref{disc}, we make comcluding statements in Section \ref{concl}.

\section{Methods in Effective Field Theory (EFT)} \label{sec2}

\subsection{Generalizing Feynman Rules in Q/EFT}

In any quantum or effective field theory, there are Feynman rules that give lines and vertices distinct mathematical factors and operators. Recalling Richard Feynman's anecdotal thought experiment to generalize the single barrier, double slit experiment into an ``infinite barrier, infinite slit experiment", quantum particles that obey wave-particle duality must follow all possible, random-step paths between two position points, with the different amplitudes under different phases $\mathcal{A}=\exp(i\varphi)$ summed linearly. 

For a single wave-particle in a semi-classical perspective (i.e. using the de Broglie formula $\lambda=h/mv$ for the wavelength, the Planck-Einstein relation $f=E/h$ for the frequency, and $E=mv^2/2+V$ for the total energy) \cite{Feynman:1948ur}, the phase of the wave with time- and position-dependent parts follows as a dimensionless action integral, with dimensionality canceled out by the Planck constant $\hbar$:
\begin{equation}
\varphi\equiv \sum_{\Delta t\rightarrow0}\left(\frac{2\pi}{\lambda}\frac{\Delta x}{\Delta t}-2\pi f \right)\Delta t\quad\Rightarrow\quad\frac{1}{\hbar}S.
\end{equation}
Here, $S=\int \mathcal{L} dt$ classically, with $\mathcal{L}=mv^2/2-V$ being the classical Lagrangian and $dx/dt=v$ implied above. If one were to generalize this derivation, such that the Lagrangian $\mathcal{L}$ can hold any form and the action $S$ can be written as its general-relativistic revision:
\begin{equation}
S=\int \mathcal{L}(x^\mu,\dot{x}^\mu;\tau)\sqrt{-g}\,d^4x,
\end{equation}
then we may apply this to relativistic particles, whose wavefunctions smear into continuous fields analogous to worldlines. The more Lagrangians we add for additional particles and their shared interactions, e.g. to describe an elastic two-particle collision, the action expands linearly and the amplitude becomes a product:
\begin{equation}
S=\sum_j S_j,\quad\Rightarrow\quad \mathcal{A}=\prod_j \exp\left(\frac{i}{\hbar}S_j\right).
\end{equation}
The linearization of amplitudes $\mathcal{A}$ comes from the analysis of different diagrams of the same event, typically higher order corrections. In this work, we imply Local Minkowski Coordinates to enforce local flatness. This choice is made in the logic that Earth-based GW detection by e.g. LIGO abhers to a flat background, similar to particle collision events captured in e.g. the Large Hadron Collider at CERN. In addition, the Minkowski metric with $\sqrt{-g}=1$ enables us to use conventional QFT principles to approach our problem.

\begin{figure}[h!]
\centering
	\subfigure[]{\includegraphics[width=0.25\textwidth]{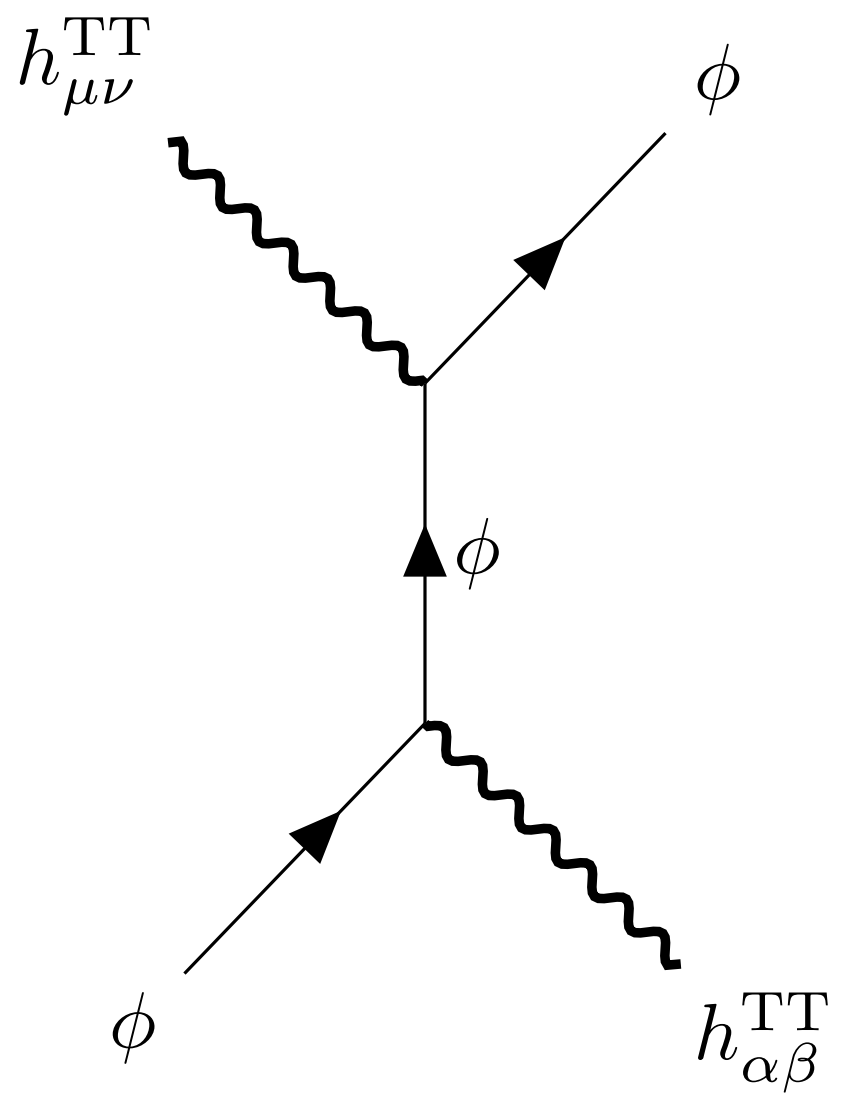}\label{fig:feynt}}
	\hspace{2cm}
	\subfigure[]{\includegraphics[width=0.35\textwidth]{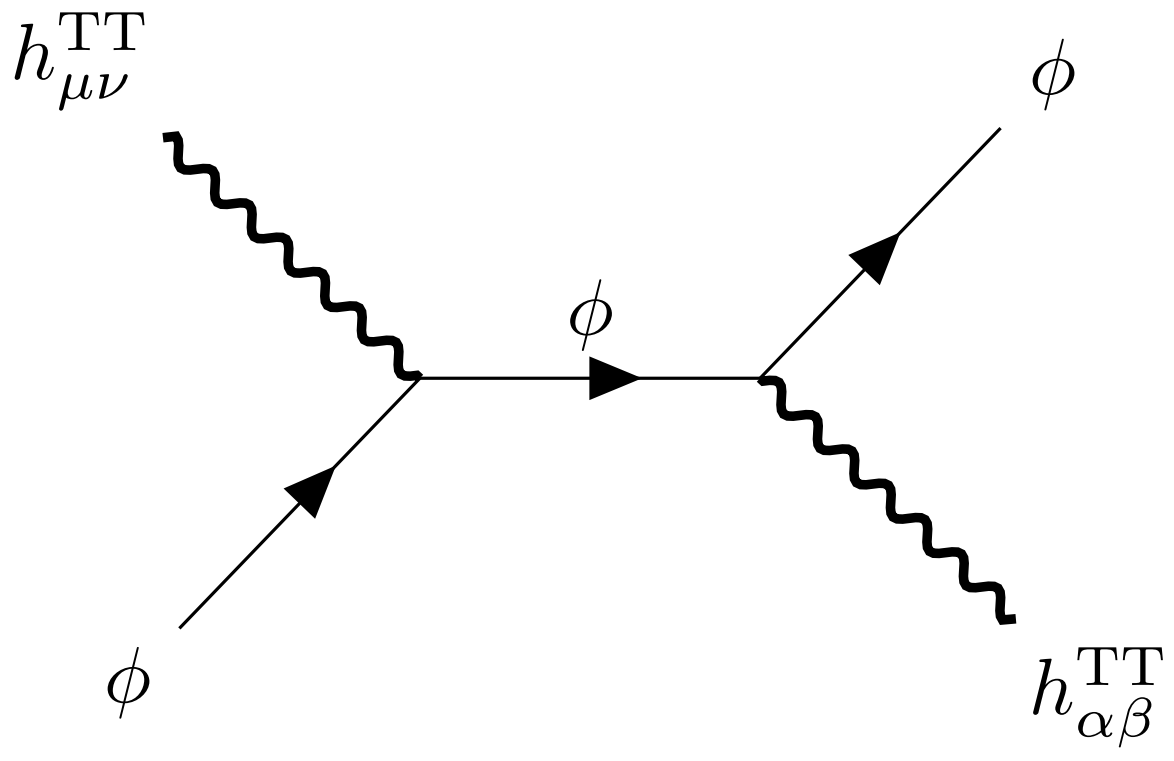}\label{fig:feyns}}
\caption{\label{fig:feyn} EFT Feynman diagrams of scalar-graviton Compton scattering. (a) a momentum-transfer ``$t$-channel" diagram; (b) a center of momentum ``$s$-channel" diagram. Arrows along the scalar lines intend to show momentum flow.}
\end{figure}

Pictorally, collision events are drawn by Feynman diagrams: at basic tree level, the external and internal lines conjoined by vertices are mathematical multiples that calculate the amplitude $\mathcal{A}$ of the corresponding event. Any event may be pictorally expressed beyond the basic diagram configuration, or as a combinatoric of the original diagram. The total amplitude of that event is calculated by the sum of the individual amplitudes of each diagram. An example of Feynman diagrams appear in Figure \ref{fig:feyn}, which happens to be the diagrams we will compute in this work.

More relevant to the diagrams we aim to calculate, the GW-graviton (for consistency under the TT-gauge) is represented as the thick wavy line. The scalar lines, drawn as the solid lines with arrows to track the flow of momentum -- given a horizontal arrow of time --, represent the suspended mass that is otherwise at rest in the laboratory (lab) frame.

 Every element to a Feynman diagram serves a distinct role:
\begin{enumerate}
\item \textbf{External Lines}: resembling the particles coming in and going out of an interaction, external lines encode asymptotic states. The factors associated with these states are given by on-shell solutions of each free particle's equation of motion, with the factors formulated by the Lehmann-Symanzik-Zimmermann (LSZ) reduction method. Examples, including the most relevant ones for our analysis, include:
\begin{itemize}
\item Scalar fields ($\phi$): 1,
\item Spin-1/2 fermions ($\psi$): $u(p),~\bar{u}(p)$ (Note: (anti-)quarks also have color states),
\item Spin-1 bosons ($A^\mu$): $\varepsilon^\mu(p)$ (Note: gluons also have color states),
\item Spin-2 bosons ($h_{\mu\nu}$): $\varepsilon_{\mu\nu}(p)$.
\end{itemize}
A particle's equation of motion is derived, provided its Lagrangian $\mathcal{L}$, via the Euler-Lagrange equation differentiated with respect to the particle field:
\begin{equation}
\frac{\partial\mathcal{L}}{\partial \Psi^{a_1\dots a_m}}-\partial_\mu\left(\frac{\partial\mathcal{L}}{\partial (\partial_\mu \Psi^{a_1\dots a_m})} \right)=0,
\end{equation}
where $\Psi^{a_1\dots a_m}$ is a tensor particle field of rank $m$. In other cases, equations of motion are derived separately (e.g. the Dirac equation for fermions), from which the respective Lagrangians may be constructed in an inverted, ``reverse-step" derivation. 

\item \textbf{Propagators}: resembling the internal particle lines between any two vertices, propagator factors are obtained by the free particle's Lagrangian, provided the general form as
\begin{equation}
\mathcal{L}=\Psi_{a_1\dots a_m}\hat{O}\Psi^{a_1\dots a_m}.
\end{equation}
Here, $\hat{O}$ is some operator in physical space. One then Fourier-transforms this operator to momentum space, whereby quantization turns 4D spacetime differential operators into momentum 4-vectors via $\partial^\mu\rightarrow -ip^\mu$. This defines the inverse element $i\tilde{D}^{-1}$ with the physical-space operator transformed into a momentum factor $P$ containing the 4-momentum contributions. The propagator is defined, therefore, by inverting the momentum-space inverse-element:
\begin{equation}
\tilde{D}=\frac{i}{P}.
\end{equation}
The inversion encodes the relevant Green's function and determines how internal momentum flows along the internal lines.

\item \textbf{Vertices}: resembling the points of interaction between external and internal lines, vertex factors are obtained by the interaction part of the action,
\begin{equation}
S_\mathrm{int}=\int \mathcal{L}_\mathrm{int}d^4x.
\end{equation}
In perturbation theory, one expands the action by expanding $\mathcal{L}_\mathrm{int}$, such that the amplitude $\exp({iS_\mathrm{int}/\hbar})$ becomes a product. Therefore, each additive term in $S_\mathrm{int}$ becomes a multiple factor, which itself constructs the vertex factor. The rule follows:
\begin{enumerate}
\item introduce a factor of $i$,
\item include the coupling constant $\kappa$,
\item insert any symmetry/combinatorial factors (following the tensor rank of the gauge boson) $C^{a_1\dots a_m}$,
\item impose momentum conservation via the 4D Dirac delta:
\end{enumerate}
\begin{equation}
\bullet\,= i\kappa \times (2\pi)^4 \delta^4\left(\sum p_\mathrm{in}-\sum p_\mathrm{out}\right)C^{a_1\dots a_m}.
\end{equation}
In typical QFTs, the gauge coupling is often defined as $g$ (not to be confused with the metric determinant) that is either treated as a constant value (e.g. $g\propto1/\sqrt{137}$ in QED) or as a dynamically running coupling (e.g. asymptotic freedom in QCD). The gauge coupling, in the case of gravitational EFT, is typically of 0.5PM order $\kappa\sim\sqrt{G}$. Thus, the PM order $m$ is gauged by $G^m\sim\kappa^{2m}$, following conventional PM expansion.
\end{enumerate}

\noindent From here on, we set $\hbar=c=1$. Thus, when refering to the graviton's energy, $E_\mathrm{graviton}\equiv \hbar\omega_\mathrm{GW}\rightarrow\omega_\mathrm{GW}$.

\subsection{Quick Comments}

The variables $t$ and $s$ refer to the standard Lorentz-invariant Mandelstam variables:
\begin{equation}
s=(p^\mu_1+p^\mu_2)^2=(p^\mu_3+p^\mu_4)^2,\quad t=(p^\mu_1-p^\mu_3)^2=(p^\mu_2-p^\mu_4)^2,
\end{equation}
where $s=E_\mathrm{CM}^2$ relates to the square of the center-of-momentum (CM) energy often associated with head-on collisions, and $t$ relates to the momentum transfer often associated with deflection/repulsion. In linear combination with the third Mandelstam variable $u=(p^\mu_1-p^\mu_4)^2=(p^\mu_2-p^\mu_3)^2$ describing another mode of momentum transfer, the three variables satisfy the total square-mass relation:
\begin{equation}
s+t+u=\sum_{i=1}^4 m^2_i.
\end{equation}
From this point on, unless otherwise specified, the spacetime indices for the 4-momenta are omitted for simplicity. The 4-momentum is henceforth labeled as $p_i$, and the 3-momentum component is denoted as $\vec{p}_i$, with the momentum indices $i$ refering to the incoming/outgoing status of the particles. I.e., in reference to Figure \ref{fig:feyn}, the GW/graviton has incoming 4-momentum $p_1$ and outgoing 4-momentum $p_3$ (having the corresponding 3-momenta) and the scalar particle has incoming 4-momentum $p_2$ and outgoing 4-momentum $p_4$ (having the corresponding 3-momenta).

For a LIGO-like Compton scattering event, the only mass present is that of the suspended mirror in laser interferometer observatories. GW-gravitons, albeit coherently having energies at the solar-mass scale, i.e. $E_\mathrm{GW}\equiv NE_\mathrm{g}\sim0.1-10 M_\odot$, are massless. Therefore, $s+t+u=2m_M^2$, where $m_M$ is the resting mirror mass. For the individual variables, we can freely use the momenta in the lab frame, in which a propagating GW-graviton with quantum energy $E_\mathrm{g}=\omega_\mathrm{GW}$ and 3-momentum $|\vec{p}|=E_\mathrm{g}$ meets and distorts the hanging mirror of mass-energy $E_M=m_M$ that is at rest in a suspended state $|\vec{p}_M|=0$:
\numparts
\begin{eqnarray}\label{sgw}
&s=p_1^2+p_2^2+2p_1\cdot p_2=m_M\left(m_M+2\omega_\mathrm{GW}\right),\\\label{tgw}
&t=p_1^2+p_3^2-2p_1\cdot p_3 = 0,\\\label{ugw}
&u=p_1^2+p_4^2-2p_1\cdot p_4 = m_M\left(m_M-2\omega_\mathrm{GW}\right).
\end{eqnarray}
\endnumparts
 We remind that we use the metric signature $\eta_{\mu\nu}=\mathrm{diag}(+,-,-,-)$, which is rather unconventional in GR but widely used in particle physics and Q/EFT. 

We essentially assume here that (in the lab frame) both the GW energy and the rest mass-energy of the mirror remain the same as prior to their collision, see footnote\footnote{In the quest to detect gravitons from incoming GWs, this may lie in the direction of measuring any slight deviation in the GW energy as a form of dispersion lost. This would then be transfered to the hanging mirror, inducing some form of kinetic energy. This is only hypothetical, as proposing means to detect gravitons is beyond the scope of this study.}. The relations in Eqs. (\ref{sgw})--(\ref{ugw}) are retrievable under the alternate definitions for $s,t,u$. Through the sum $s+t+u$, the known result $2m_M^2$ is recovered. More intriguing is the relation of the CM energy-squared, directly defined as Eq. (\ref{sgw}), in terms of the lab frame energies:
\begin{equation}\label{cmeng}
E_\mathrm{CM}^2=m^2_M+2m_M\omega_\mathrm{GW}.
\end{equation}
The first term is essentially the rest energy-squared of the suspended mass, and the second term $2m_M\omega_\mathrm{GW}$ is the sub-leading contribution that gauges the interaction. Therefore, considering the second term for the CM energy leads to the calculation $ E_\mathrm{CM}=\sqrt{2m_M\omega_\mathrm{GW}}\sim 10^{10.5}$ MeV (or equivalently $10^{1.5}\simeq31.6$ PeV, for a suspended mirror mass of $m_M\sim40$ kg and the peak frequency $\omega \sim 100$ Hz). In the field of astroparticle physics, the PeV energy scale signifies a threshold for extreme, ultra-high-energetic phenomena within our galaxy. Having $\sim32$ PeV as the CM energy between the suspended mass and the GW graviton is uniquely beyond the infrared analyses previously considered for gravitational Compton scattering. Furthermore, if gravitons were physical entities beyond useful analytical tools in EFT, their emissions from quadrupolar sources as part of the larger coherent emission of GWs may be classified as one of the most extreme cases of astroparticle physical phenomenon.

\subsection{Governing Equations and Gauges} \label{feynrules}

Analogous to electrodynamic Compton scattering, LIGO-like Compton scattering between a test mass and an incoming graviton can be drawn by the diagrams readily provided in Figure \ref{fig:feyn}. The Lagrangian of a Compton scattering event between a scalar particle and a graviton contains the individidual Lagrangians for the scalar particle and the graviton, as well as the Einstein-Hilbert Lagrangian gauging the interaction:
\begin{equation}
\mathcal{L}=\mathcal{L}_\mathrm{Scalar}+\mathcal{L}_\mathrm{Graviton}+\mathcal{L}_{\mathrm{EH}}.
\end{equation} 
The tree-level Feynman rules for scalar-graviton and graviton-graviton scattering diagrams are provided in Refs. \cite{Feynman:2002, Rafie-Zinedine:2018izq}. However, for Compton scattering, the relevant diagrams in the CM frame include the $t$- and $s$-channel diagrams shown in Figure \ref{fig:feyn}. Therefore, we only need to know the terms associated with scalar external lines and propagator, graviton external lines, and scalar-graviton vertices.

\subsubsection{Scalar External Lines and Propagator}

In Minkowski spacetime, scalar particles governed by the field $\phi$ follow the Lagrangian
\begin{equation} \label{scallag}
\mathcal{L}_{\mathrm{Scalar}}=\frac{1}{2}(\partial_\mu\phi)(\partial^\mu\phi)-\frac{1}{2}m^2\phi^2,
\end{equation}
where $m$ is the mass of the scalar particle, e.g. $m_M$. Applying this Lagrangian to the Euler-Lagrange equation yields the Klein-Gordon equation as the corresponding equation of motion:
\begin{equation}
(\partial_\mu\partial^\mu-m^2)\phi=0,
\end{equation}
where $\partial_\mu \partial^\mu\equiv\eta_{\mu\nu}\partial^\mu\partial^\nu$ defines the d'Alembert wave operator $\Box\equiv\partial^2_t-\nabla^2$ under the metric signature $\eta_{\mu\nu}=\mathrm{diag}(+,-,-,-)$. This scalar field has a plane wave solution that satisfies orthonormality $\langle\phi_i|\phi_j\rangle=\delta_{ij}$. Thus, external scalar lines are represented by their LSZ reductions of unitary value $1$.

Because the scalar particle serves the role as the internal particle in both diagrams offered in Figure \ref{fig:feyn}, one reorganizes Eq. (\ref{scallag}) in the form $\mathcal{L}=\phi\hat{O}\phi$, such that inversion and Fourier-space quantization is possible:
\begin{equation}
\begin{tikzpicture}
  \begin{feynman}
    \vertex (a);
    \vertex [left=2cm of a] (f1);

    \diagram* {
      (a) -- [anti fermion, thick] (f1),
    };
  \end{feynman} 
\end{tikzpicture}~=~\frac{i}{q^2-m^2}.
\end{equation}
where $q$ is the internal 4-momentum between the vertices.

\subsubsection{Graviton External Lines}

As provided in Ref. \cite{Rafie-Zinedine:2018izq}, the Lagrangian of a graviton with a tensor field $h_{\mu\nu}$ is given by
\begin{equation}
\mathcal{L}_{\mathrm{Graviton}}=\frac{1}{2}(\partial_\alpha h^{\mu\nu})(\partial^\alpha h_{\mu\nu})-\frac{1}{4}(\partial_\alpha h_\mu^{~\mu})(\partial^\alpha h_{\nu}^{~\nu}).
\end{equation} 
The first term describes the kinetic energy, and the second term ensures gauge invariance. Since $h_{\mu\nu}$ is a rank-2 tensor, the graviton has spin $s=2$. A massless graviton, however, has only two degrees of freedom (d.o.f); this corresponds to the two orthogonal polarization states of a propagating GW. Thus, massless gravitons have the traceless and transverse (TT) gauge that GWs have. For TT-gauged gravitons with the field $h^{\mathrm{TT}}_{\mu\nu}$, the traceless condition yields $h_\mu^{~\mu}=h_\nu^{~\nu}=0$. Simplifying terms involving $h_{\mu\nu}$ under this TT-gauge and neglecting second-order perturbations $\mathcal{O}(h^2)$, the free wave equation and its plane wave solution are
\begin{equation} \label{graveq}
\Box h^{\mathrm{TT}}_{\mu\nu}=0~~~\rightarrow~~~h^{\mathrm{TT}}_{\mu\nu}(x)\propto\exp\left(-ip_\rho x^\rho \right)\varepsilon^{\mathrm{TT}}_{\mu\nu}(p).
\end{equation}
This essentially recovers Eq. (\ref{gwvacuum}). Here, however, the 4-vector for the wave number $k^\mu$ is replaced with the 4-momentum $p^\mu=(E,\vec{p}\,)$ with an effective Planck constant scaling, and $\varepsilon^\mathrm{TT}_{\mu\nu}(p)$ is the TT-gauged polarization tensor:
 \begin{equation}
\varepsilon_{\mu\nu}^{\mathrm{TT}}(p)=\left( \begin{array}{cccc} 
0&0&0&0\\
0&h_+&h_\times&0\\
0&h_\times&-h_+&0\\
0&0&0&0 \end{array} \right).
\end{equation}
Furthermore, the TT-gauge sets the conditions $\varepsilon_{\mu\nu}^{\mathrm{TT}}\eta^{\mu\nu}=0$ for tracelessness and $\varepsilon_{\mu\nu}^{\mathrm{TT}}p^\mu=\varepsilon_{\mu\nu}^{\mathrm{TT}}p^\nu=0$ for transverse motion. TT-gauge properties will prove to be instrumental in our analysis. As external particle lines, TT-gauged gravitons are described by external polarization tensors $\varepsilon_{\mu\nu}^{\mathrm{TT}}(p_i)$ entering the vertex and $\varepsilon_{\mu\nu}^{*~\mathrm{TT}}(p_j)$ exiting the vertex. 

\subsubsection{Scalar-Graviton Vertex}

 For a gravitational field expressed by the EFEs, the Einstein-Hilbert Lagrangian governs the dynamics:
\begin{equation} \label{gravlag}
\mathcal{L}_{\mathrm{EH}}=-\frac{1}{16\pi G}R,
\end{equation}
where $R\equiv g_{\mu\nu}R^{\mu\nu}$ is the Ricci scalar, the trace of the Ricci curvature tensor. In flat spacetime ($g_{\mu\nu}=\eta_{\mu\nu}$), $\mathcal{L}_{\mathrm{EH}}=0$ since $R=0$. However, given a linearized metric $g_{\mu\nu}=\eta_{\mu\nu}+\kappa^2 h_{\mu\nu}$ with inverse $g^{\mu\nu}=\eta^{\mu\nu}-\kappa^2 h^{\mu\nu}$ and gauge coupling $\kappa$, the Ricci scalar depends on the pertubation: $R=\kappa^2 h_{\mu\nu}R^{\mu\nu}=\delta R$, defining a small deviation in the Lagrangian $\delta\mathcal{L}$. It is this deviated Lagrangian that will gauge the interaction.

The graviton sector in the interaction is the gauged metric $\kappa^2 h_{\mu\nu}$, and the scalar sector rests in the Ricci tensor. More specifically, we utilize the alternative expression of the EFEs, where $R^{\mu\nu}=8\pi G(T^{\mu\nu}-Tg^{\mu\nu}/2)$, applying the linearized inverse metric for $g^{\mu\nu}$. Readily considering the traceless gauge and omitting self-contracting terms $\mathcal{O}(h_{\mu\nu}h^{\mu\nu})$, we have 
\begin{equation} \label{gravlag2}
\delta\mathcal{L}=-\frac{\kappa^2}{2}h_{\mu\nu}T^{\mu\nu},
\end{equation}
where the energy-momentum tensor is that of the scalar field on the flat background:
\begin{equation}
T^{\mu\nu}=\frac{1}{m}\left[\partial^{(\mu} \phi\, \partial^{\nu)}\phi - \eta^{\mu\nu}\left(\partial_\beta \phi \partial^\beta \phi + m^2\phi^2 \right)\right].
\end{equation}
Here, $\partial^{(\mu} \phi \,\partial^{\nu)}\phi =\partial^{\mu}\phi \,\partial^{\nu}\phi+\partial^{\nu}\phi \,\partial^{\mu}\phi$ is a symmetric summation. The Lagrangian of the interaction is presented in the form $\mathcal{L}=\phi\hat{O}\phi$ mainly via the energy momentum tensor. It is therefore possible to take the physical space contribution and Fourier-transform it into the momentum space. 

\begin{figure}[h!]
\centering
\includegraphics[width=0.25\textwidth]{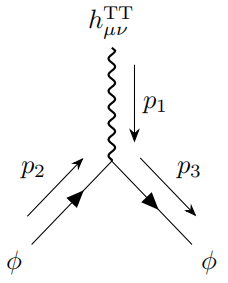}
\caption{\label{fig:vertex} A scalar-graviton vertex.}
\end{figure}

While the symmetrized spacetime derivatives appear redundant at the level of the Lagrangian, they correspond to distinct momentum contributions at the scalar-graviton vertex, depicted in Figure \ref{fig:vertex}. Each derivative acts on a different scalar field, corresponding to the momentum of the scalar line it acts on. For the given flow of the 4-momenta in Figure \ref{fig:vertex}, the vertex factor is given as follows, cf. Refs. \cite{Feynman:2002, Rafie-Zinedine:2018izq}:
\begin{equation}\label{sgvert}
V^{\mu\nu}=i\kappa\left[p_2^{(\mu} p_3^{\nu)}-\eta^{\mu\nu}(p_2\cdot p_3-m^2) \right]\times(2\pi)^4\delta^4(p_1+p_2-p_3).
\end{equation} 

Here, the spacetime indices of the vertex tensor couple with those of the TT-polarization tensor. The vertex factor depends solely on the scalar particle's 4-momenta explicitly, with the graviton's 4-momentum implied in the term containing the metric $\eta^{\mu\nu}$. If the direction of any scalar line's 4-momentum is flipped, a negative sign is introduced to that respective momentum in the square-bracket term (but not inside the Dirac delta). Similarly, reversing the graviton's momentum only flips the sign on metric term $\eta^{\mu\nu}$. 

Given this is a deviated Lagrangian, sourced by the interaction, we refer to the GW equation via the linearized EFEs with a non-zero source term:
\begin{equation}
\Box\bar{h}^{\mu\nu}=-16\pi G T^{\mu\nu}.
\end{equation}
After one contracts both sides with the GW metric $h_{\mu\nu}$, such that the left-hand side (in 4-momentum space) defines the deviated Lagrangian and the right hand side matches that of Eq. (\ref{gravlag2}), we define readily the gauge coupling of the interaction to be $\kappa^2=32\pi G$. This is the gauge coupling we use in this work; for completeness, other literature use e.g. $\kappa^2=8\pi G$ \cite{Rafie-Zinedine:2018izq}, or equivalently expressed in terms of the Planck mass $\kappa^2=8\pi/m_P^2$ \cite{Delgado:2022uzu, Herrero-Valea:2022lfd} with $m_P^2=\hbar/G$ (briefly recovering $\hbar$).

\section{Calculating $\mathcal{A}_{\mathrm{tot}}$} \label{scatt}

Using the Feynman rules in Section \ref{feynrules}, the total amplitude, $\mathcal{A}_{\mathrm{tot}}=\mathcal{A}_t+\mathcal{A}_s$, is determined using Feynman calculus. The standard rulebook to follow Feynman calculus can be found in relevant chapters of Ref. \cite{griffiths}. This amplitude describes, in the CM frame, the hypothetical scattering of a scalar mass by a single TT-gauged graviton (otherwise part of the coherent-state bulk of the GW). To connect this result with observable phenomena, a transformation into the lab frame --  the frame where GWs are detected -- and relevant coherence state occupation enhancment is necessary. 

\subsection{$t$-channel Compton scattering}

After following the Feynman rules, the amplitude calculation for the $t$-channel in Figure \ref{fig:feynt} is 
\begin{eqnarray}
\mathcal{A}_t=&\,\frac{\kappa^2}{t-m_M^2}\\\nonumber
&\times \varepsilon^{*~\mathrm{TT}}_{\alpha\beta}(p_4)\left[(p_2^\alpha (p_2^\beta-p_4^\beta)+(p_2^\alpha-p_4^\alpha) p_2^\beta)+\eta^{\alpha\beta}(p_2\cdot (p_2-p_4)) \right]\\\nonumber
&\times \varepsilon^{\mathrm{TT}}_{\mu\nu}(p_1)\left[((p_2^\mu-p_4^\mu) p_3^\nu+p_3^\mu (p_2^\nu-p_4^\nu))-\eta^{\mu\nu}((p_2-p_4)\cdot p_3) \right].
\end{eqnarray}
Here, $t=(p_2-p_4)^2=(p_1-p_3)^2$ is the Lorentz-invariant momentum transfer. The internal momentum is chosen to flow through the bottom vertex upward. Applying the TT gauge to the graviton polarization tensor greatly simplifies the expression:
\begin{equation} \label{simpt1}
\mathcal{A}_t=\frac{2\kappa^2}{t-m_M^2}~\varepsilon^{*~\mathrm{TT}}_{\alpha\beta}(p_4)p_2^\alpha p_2^\beta\times \varepsilon^{\mathrm{TT}}_{\mu\nu}(p_1)\left[p_3^\nu(p_2^\mu-p_4^\mu) +p_3^\mu (p_2^\nu-p_4^\nu) \right].
\end{equation}

Because the tensor elements are real, $\varepsilon^{*~\mathrm{TT}}_{\alpha\beta}=\varepsilon^{\mathrm{TT}}_{\alpha\beta}$. The contraction of a TT-polarization tensor under the 4-momentum vector $p_b$ with a different 4-momentum vector $p_a$ is given by
\begin{equation} \label{polmom}
\varepsilon^{\mathrm{TT}}_{\alpha\beta}(p_b)p_a^\alpha=\Big[0, ~ h^+_bp_{ax}+h^\times_{b}p_{ay}, ~ h^\times_bp_{ax}-h^+_bp_{ay}, ~ 0 \Big],
\end{equation}
subsituting this into the simplifed amplitude yields (with the prime notation $^\prime$ denoting the momenta and polarizations under the CM frame)
\begin{eqnarray} \label{simpt2}
\mathcal{A}_t=&\frac{4\kappa^2}{t-m_M^2}\Big[\left(h^+_4(p^{'\,2}_{2y}-p^{'\,2}_{2x})-2h^\times_{4}p'_{2x}p'_{2y}\right)\\\nonumber
&\times \left((h^+_1p'_{3x}+h^\times_{1}p'_{3y})(p'_{4x}-p'_{2x})+(h^\times_1p'_{3x}-h^+_1p'_{3y})(p'_{4y}-p'_{2y}) \right)\Big].
\end{eqnarray}
We begin to see that the amplitude depends on both incoming and outgoing TT-polarization states and the 3-momentum components of both the scalar particle and the outgoing graviton.

In the CM frame, the incoming and outgoing 3-momenta are equal in magnitude but opposite in direction, reflecting the symmetry of the collision: $\vec{p}\,'_{1,3}=-\vec{p}\,'_{2,4}$. The scattering from an incoming particle into an outgoing one is described by the scatting angle $\theta$, such that $\vec{p}\,'_{1,2}=\vec{p}\,'_{3,4}\cos\theta$. Applying these relations to individual components, i.e., $p'_{1i,2i}=p'_{3i,4i}\cos\theta$ and $p'_{1i,3i}=-p'_{2i,4i}$, allows the simplification of Eq. (\ref{simpt2}) in terms of only one 3-momentum and its components, chosen here as the outgoing scalar particle momentum $\vec{p}\,'_3$:
\begin{eqnarray} \label{simpt3}
\mathcal{A}_t=&\frac{4\kappa^2}{t-m_M^2}\cos^2\theta(1-\cos\theta)\\\nonumber
&\times \Big[h^+_4h^+_1(p^{'\,2}_{3y}-p^{'\,2}_{3x})^2+4h^\times_{4}h^\times_1p^{'\,2}_{3x}p^{'\,2}_{3y}\\\nonumber
&~~~~~~-  2(p^{'\,2}_{3y}-p^{'\,2}_{3x})p'_{3x}p'_{3y}\left(h^+_4h^\times_1+h^\times_4h^+_1 \right)\Big].
\end{eqnarray}
In the lab frame, this momentum corresponds to the hanging mass' recoil motion after its interaction with the incoming GW. In the CM frame, the relation $\cos\theta=1+2st/(s-m_M^2)^2$ applies, and the magnitude of the 3-momentum is related to $s$ via $|\vec{p}\,'_3|^2=(s-m_M^2)^2/(4s)$.

The existing x- and y-components of $\vec{p}\,'_3$ construct the transverse momentum $\vec{p}\,'_T$, expressed in polar coordinates as trigometric functions of a transverse-plane angle $\psi$: $p'_{3x}=|\vec{p}\,'_T|\cos\psi$ and $p'_{3y}=|\vec{p}\,'_T|\sin\psi$. Since the angular range on the transverse plane $\psi\in[0,\,2\pi]$  is unaffected by the plane's inclination via the scattering angle $\theta$, integration over all transverse angles is applied to the x- and y-components of $\vec{p}\,'_3$: 
\numparts
\begin{eqnarray}
&(p^{'\,2}_{3y}-p^{'\,2}_{3x})^2=|\vec{p}\,'_T|^4 \int_0^{2\pi}(\sin^2\psi-\cos^2\psi)d\psi=\pi |\vec{p}\,'_T|^4,\\
&p^{'\,2}_{3x}p^{'\,2}_{3y}=|\vec{p}\,'_T|^4\int_0^{2\pi}(\cos^{2}\psi\,\sin^{2}\psi)d\psi =\frac{\pi}{4}|\vec{p}\,'_T|^4,\\\nonumber
& (p^{'\,2}_{3y}-p^{'\,2}_{3x})p'_{3x}p'_{3y}=|\vec{p}\,'_T|^4\int_0^{2\pi}\left( (\sin^{2}\psi-\cos^{2}\psi)\cos\psi\,\sin\psi\right)d\psi\\
&\quad\quad\quad\quad\quad\quad\quad\,=0.
\end{eqnarray}
\endnumparts
With $|\vec{p}\,'_T|=|\vec{p}\,'_3|\sin\theta$, the transverse momentum can be rewritten in terms of $t$ and $s$ via $\sin\theta=\sqrt{1-\cos^2\theta}$ and the Lorentz-invariant definitions of $\cos\theta$ and $|\vec{p}\,'_3|$. Consequently, the $t$-channel amplitude in the CM frame is 
\begin{eqnarray} \label{simpt4}
\mathcal{A}_t=&-\frac{\pi\kappa^2}{2}\frac{t}{t-m_M^2}\left(\frac{s}{(s-m_M^2)^2}\right)^{-1}\left(1+\frac{2st}{(s-m_M^2)^2}\right)^2\\\nonumber
&\times \left(\frac{4st}{(s-m_M^2)^2}+\frac{4s^2t^2}{(s-m_M^2)^4}\right)^2 \Big[h^+_4h^+_1+h^\times_{4}h^\times_1 \Big].
\end{eqnarray}

\subsection{$s$-channel Compton scattering}

After following the Feynman rules and applying the TT-gauge, the $s$-channel amplitude shown as Figure \ref{fig:feyns} is given by
\begin{eqnarray}
\mathcal{A}_s=\frac{2\kappa^2}{s-m_M^2}&~\varepsilon^{\mathrm{TT}}_{\alpha\beta}(p_4)p_3^\alpha p_3^\beta\\\nonumber
&\times \varepsilon^{\mathrm{TT}}_{\mu\nu}(p_1)\left[(p_2^\mu (p_3+p_4)^\nu+p_2^\nu(p_3+p_4)^\mu)\right].
\end{eqnarray}

Using the relation in Eq. (\ref{polmom}) and applying the CM condition $p'_{3i}=-p'_{4i}$, the $s$-channel amplitude simplfies to a null result:
\begin{eqnarray}
\mathcal{A}_s&=\frac{4\kappa^2}{s-m_M^2}\Big[\left(h^+_4(p^{'\,2}_{3y}-p^{'\,2}_{3x})-2h^\times_{4}p'_{3x}p'_{3y}\right)\\\nonumber
&\quad\times \left((h^+_1p'_{2x}+h^\times_1p'_{2y})(-p'_{3x}-p'_{4x})+(h^\times_1p'_{3x}-h^+_1p'_{3y})(-p'_{3y}-p'_{4y}) \right)\Big]\\\nonumber
&=0.
\end{eqnarray}

\subsection{Total Amplitude} \label{tamp}

Therefore, the total amplitude of scalar-graviton Compton scattering is solely governed by the $t$-channel amplitude. As it is primarily expressed in terms of Lorentz-invariant variables, the transformation into the lab frame only affects the configuration of the plus and cross polarizations. Specifically, the polarizations transform as $h_{1,4}^{+,\times}\rightarrow h_{+,\times}$, as long as there is non-existent (or negligibly small) dispersion in the lab frame, i.e. in a LKV detection event:
\begin{eqnarray} \label{labfrm}
\mathcal{A}_\mathrm{Lab}=&-\frac{\pi\kappa^2}{2}\frac{t}{t-m_M^2}\left(\frac{s}{(s-m_M^2)^2}\right)^{-1}\left(1+\frac{2st}{(s-m_M^2)^2}\right)^2\\\nonumber
&\times \left(\frac{4st}{(s-m_M^2)^2}+\frac{4s^2t^2}{(s-m_M^2)^4}\right)^2 \Big[h^2_++h^2_\times \Big].
\end{eqnarray}
Here, one may recognize $\left[h_+^2+h_\times^2\right]$, in the context of GW astrophysics, as the invariant polarization sum that is proportional to the GW energy (flux) density. Knowing that these polarizations are those of the single graviton, a coherent enhancement recovers the classical configuration that one can define from a macroscopic waveform model of choice, either numerical or (semi-)analytical.

At this stage, it is a standard EFT practice to discard the graviton polarizations from Eq. (\ref{labfrm}), while preserving only the kinematic information of the amplitude. E.g., the WQFT framework does this in order to retain effective end-state measurables in the PN regime of coalescence. However, as we describe the graviton-matter interaction that is omnipresent in a GW-mirror interaction, one may find removing the graviton sector counterintuitive. As a heuristics exercise, we keep the polarizations, and we discuss the ramifications of leaving the polarizations in the amplitude. 

After computing the magnitude squared of Eq. (\ref{labfrm}), defined as $|\mathcal{A}_\mathrm{Lab}|^2=\mathcal{A}^*_\mathrm{Lab}\mathcal{A}_\mathrm{Lab}$, one can determine the differential cross section with respect to $t$ via the expression \cite{griffiths}:
\begin{equation}
\frac{d\sigma}{dt}=\frac{1}{16\pi s^2}|\mathcal{A}_\mathrm{Lab}|^2.
\end{equation}
This computes the total cross section $\sigma\equiv\int d\sigma$ over the range of $t\in[-s,0]$. After evaluating the integral, we readily consider $s\equiv E^2_\mathrm{CM}=2m_M\omega_\mathrm{GW}$ for the CM energy-squared, while also implying a Taylor expansion for small inverse-mass $m_M^{-1}$ to yield the leading-order term in the cross section:
\begin{eqnarray}
\sigma&=\left.\int_{-s}^0 \frac{dt}{16\pi s^2}\mathcal{A}^*_\mathrm{Lab}\mathcal{A}_\mathrm{Lab}\right|_{s\rightarrow E^2_\mathrm{CM}} \\\nonumber
&\Rightarrow ~\frac{2048\pi}{5}\pi^2 G^2\left[h_+^2+h_\times^2\right]^2\frac{\omega_\mathrm{GW}^3}{m_M}+\mathcal{O}(m_M^{-2}).
\end{eqnarray}
As a defining feature of a soft-graviton, heavy-target regime, the ratio $\omega_\mathrm{GW}/m_M$, let alone the inverse mass $m_M^{-1}$, could be understood as an EFT expansion term in the total cross section. Provided $\sigma=\pi b^2$, the corresponding impact paramter $b$ is
\begin{equation}\label{gb}
 b =\sqrt{\frac{2048}{5}\frac{\omega_\mathrm{GW}}{m_M}}\pi G\left[h_+^2+h_\times^2\right]\omega_\mathrm{GW},
\end{equation}
which entails intriguing details to heuristically infer:
\begin{enumerate}
\item The rooted ratio $\mathrm{R_{root}}=\sqrt{\omega_\mathrm{GW}/m_M}\sim10^{-21}$ sets the scale of the LIGO-like scattering between a single graviton and a hanging mass. Rather interestingly, it is (by factor alone) similar to the $10^{-21}$ strain amplitude that is typical for astrophysical GWs. Recognizing that $\sqrt{\omega_\mathrm{GW}/m_M}$ scales the mirror recoil after having made contact with the GW graviton (and with it the suppression of momentum transfer), one can make a logical link between this quantity and the strain the GW applies to all objects including the hanging mirror.

\subitem One must also acknowledge that $10^{-21}$ from the rooted ratio is rather convenient, if not coincidential, provided the mass of the hanging mirror is $m_M=40$ kg and GW frequencies peak at $\sim100$ Hz in order of magnitude. 

\item The remaining terms outside the square-root: $\sim\pi G[h^2_++h^2_\times]\omega_\mathrm{GW}$, relates to a graviton-scale impact parameter whose prefactors alone scale in the order of the Planck-length-squared: $l_P^2/[\mathrm{m^2}]\sim10^{-70}$. Together with the rooted ratio scale, $[b]\sim 10^{-91}$ in order of magnitude, that is without considering the scaling orders imposed by the graviton polarizations -- if they were removed this would not be a contributing factor. 

\subitem On that note, it is fair to state that any sensible means to detect just a single graviton is beyond our current capabilities.
\end{enumerate}

\section{Discussion} \label{disc}

The calculation of the $t$-channel amplitude, provided pictorally as Figure \ref{fig:feynt}, together with the null result laid out by the $s$-channel amplitude via Figure \ref{fig:feyns}, yielded a total amplitude that recovers classical intuition of a graviton-matter interaction in the context of a LIGO-like event. More speficially, recovering the invariant sum $h^2:=h^2_+ + h^2_\times$ for the graviton polarizations enables a logical link to the classical (astrophysical) polarizations if one applies a coherent-state population enhancement. After calculating the total cross section and setting the energy scale to be the CM energy $\sqrt{s}=E_\mathrm{CM}$, the corresponding impact parameter $b$ depends on graviton-scale parameters that can be enhanced into the classical scale with proper coherent-state enhancements.

For instance, given $E_\mathrm{GW}=N\omega_\mathrm{GW}$ and supposing $h^2_\mathrm{GW}=Nh^2_\mathrm{g}$, the graviton-scale impact parameter is enhanced by $N^2$ to yield a quantity in the classical regime:
 \begin{equation}\label{gwb}
N^2 b=:b_\mathrm{C} =\sqrt{\frac{2048}{5}\frac{\omega_\mathrm{GW}}{m_M}}\pi G\left[h_+^2+h_\times^2\right]_\mathrm{GW} E_\mathrm{GW}.
\end{equation}
One may alternatively define a rescaled quantity $\tilde{b}_\mathrm{C}:=b_\mathrm{C}\sqrt{m_M/\omega_\mathrm{GW}}$, which removes the explicit dependence on the detector mass and isolates the GW sector. This yields a quantity purely dependent on astrophysical source parameters. We also approximate $\sqrt{2048/5}\approx20.24$.

Since Eq. (\ref{gwb}), and with it $\tilde{b}_\mathrm{C}=b_\mathrm{C}\sqrt{m_M/\omega_\mathrm{GW}}$, depends on the GW energy and the GW polarizations (having decided to leave them in), it is at this stage where one may use a specific waveform model of choice and its respective energy computation. In this calculation, we refer to the mass-shell model of CCBs \cite{MacKay:2024qxj, MacKay:2025uyg}, which readily provides an analytical structure of the waveforms that distinctly show the rates of change in GW frequency and binary separation across the IMR phases. This model also provides an analytical expression that approximates rather nicely the GW energy radiated at merger. Knowing that the propagating GW, when making contact with a LIGO-like detector, entails information of the merger phase within the wave's peak in amplitude and frequency, we specifically use the mass-shell model expressions for coalescence energy as $E_\mathrm{GW}$ and refer to the model's waveforms for the invariant sum $h^2_+ + h^2_\times$.

\subsection{GW profiles and energy in the mass-shell model}

For both stable classical binaries and CCBs, the reduced mass $\mu$ simplifies two-body dynamics into a singular effective system. For CCBs specifically, the binary may be assumed to behave as a singular object (i.e. a rotating and contracting hollow mass shell) when viewed from far away. While undergoing coalescence, the rate of change in the separation $-d{L}/dt$ w.r.t the observer time relates to the mass shell's contracting  diameter with the rate of change $-d{S}/dt$ w.r.t the observer time. After one integrates this equivalence over the timelapse $t'\in[t,\,t_C]$ ($t$ is dynamic and $t_C$ is a fixed coalescence time), and imposes that coalescence ends when the masses touch surfaces: $L(t_C)=r_1+r_2$, under the shell diameter that identifies the total mass horizon diameter: $S(t_C)=4GM$, we define:
\begin{equation}\label{rads}
S(t)=L(t)-r_1-r_2+4GM.
\end{equation}
In this model, CCBs are viewed instead as a shrinking mass shell with the constant measure $\mu$ and contracting radius $\rho(t)=S(t)/2$, until reaching the ``innermost'' shell that is the total mass horizon. As for the respective TT-gauged spatial waveforms, even in the CCB mass-shell model, they are conventionally quadrupolar. If we assume e.g. (nearly-)circular orbits, and demonstrate time variation in the CCB separation $S=S(t)$ and $\Omega$ while conserving angular momentum $J=I\Omega$, we eventually yield waveform expressions written entirely in terms of observer-time derivatives of e.g. $\Omega$, c.f. Refs. \cite{MacKay:2024qxj, MacKay:2025uyg}:
\numparts
 \begin{eqnarray}\label{hp2}
&h_{+}^{\mathrm{TT}}=-\frac{G\mu S^2}{D}\Big(\sin(2\Omega t)\left(-\frac{\dot{\Omega}}{\Omega}\left(\Omega +t\dot{\Omega}\right)+\dot{\Omega}+\frac{t}{2}\ddot{\Omega}\right)\\\nonumber
&\quad\quad\quad\quad\quad\quad\quad -\cos(2\Omega t)\left(-\frac{\ddot{\Omega}}{4\Omega}+\frac{\dot{\Omega}^2}{4\Omega^2}-\left(\Omega+t\dot{\Omega}\right)^2\right)-\frac{\ddot{\Omega}}{4\Omega}+\frac{\dot{\Omega}^2}{4\Omega^2}\Big),\\\label{hc2}
&h_{\times}^\mathrm{TT}=-\frac{G\mu S^2}{D}\Big(\sin(2\Omega t)\left(-\frac{\ddot{\Omega}}{4\Omega}+\frac{\dot{\Omega}^2}{4\Omega^2}-\left(\Omega+t\dot{\Omega}\right)^2\right)\\\nonumber
&\quad\quad\quad\quad\quad\quad\quad +\cos(2\Omega t)\left(-\frac{\dot{\Omega}}{\Omega}\left(\Omega +t\dot{\Omega}\right)+\dot{\Omega} +\frac{t}{2}\ddot{\Omega}\right)\Big).
\end{eqnarray}
\endnumparts
In the general sense, taking Eqs. (\ref{hp2}) and (\ref{hc2}) as the inspiral-merger basis, one may offer a complete IMR waveform by imposing a linear expansion in sine and cosine profiles, each scaled by a time-dependent, polarization-specific envelope function $\mathfrak{E}(t)$ encapsulating the dynamics:
\numparts
\begin{eqnarray} \label{piece1}
&h_{+}^\mathrm{TT}(t)=-\frac{G\mu S^2}{D}\left[\cos(2\Omega t) \mathfrak{E}_{+}^\mathrm{cos}(t)+\sin(2\Omega t)\mathfrak{E}_{+}^\mathrm{sin}(t)\right],\\\label{piece2}
&h_{\times}^\mathrm{TT}(t)=-\frac{G\mu S^2}{D}\left[\cos(2\Omega t) \mathfrak{E}_{\times}^\mathrm{cos}(t)+\sin(2\Omega t)\mathfrak{E}_{\times}^\mathrm{sin}(t)\right].
\end{eqnarray}
\endnumparts

If we equate Eq. (\ref{piece1}) to Eq. (\ref{hp2}) and Eq. (\ref{piece2}) to Eq. (\ref{hc2}) for the inspiral-merger phases, we notice uniquely that $\mathfrak{E}^\mathrm{cos}_+=-\mathfrak{E}^\mathrm{sin}_\times$ and $\mathfrak{E}^\mathrm{sin}_+=\mathfrak{E}^\mathrm{cos}_\times$. Therefore, when one finds $h^2_++h^2_\times$, one observes that all cross terms cancel out and the trigometric identity $\cos^2x+\sin^2x=1$ preserves the individual sine and cosine envelope functions. And given their symmetric relations when they are squared, the choice of referring to sine and cosine envelopes from either the cross or the plus polarization is invariant to the result:
\begin{equation}\label{hsums}
\Rightarrow\quad [h^2_+ + h^2_\times]_\mathrm{GW}=A_0^2\left[\mathfrak{E}^2_\mathrm{cos}+\mathfrak{E}^2_\mathrm{sin} \right].
\end{equation}
Here, we define $A_0:=G\mu S^2/D$ as the GW amplitude, with the frequency profile encoded in the dynamic envelope functions.

While it is conventional to extract the energy density (and energy flux density) of GWs via $h_+^2+h_\times^2$, using either both Eqs. (\ref{hp2}) and (\ref{hc2}) or Eqs. (\ref{piece1}) and (\ref{piece2}) generally, a more heuristic approach was employed in Refs. \cite{MacKay:2024qxj, MacKay:2025uyg}, wherein the relevant quantity was obtained through a variational treatment of the EFEs. In short, the Laplace-Beltrami formulation of the Ricci tensor was applied to a Kerr metric Ansatz, and the corresponding energy density $T_{00}$ of the CCB mass shell was obtained via the Einstein field equations. At the time of coalescence $t_C$, the resulting surface energy depends on the reduced mass $\mu$, the symmetric mass ratio $\alpha:=\mu/M$, and the normalized orbital spin velocity of the CCB, c.f. Ref. \cite{MacKay:2025uyg}:
\begin{equation}\label{energy}
E_\mathrm{GW}\simeq 0.826\frac{\mu^2}{M}\left(1-5.276 \beta_C^2 \right).
\end{equation}

\subsection{GW-mirror impact parameter}

With Eqs. (\ref{hsums}) and (\ref{energy}) defining model-specific waveform invariants and the coalescence energy, we obtain
 \begin{equation}
\tilde{b}_\mathrm{C} \simeq 16.72\pi GA_0^2\left[\mathfrak{E}^2_\mathrm{cos}+\mathfrak{E}^2_\mathrm{sin} \right]\frac{\mu^2}{M}\left(1-5.276 \beta_C^2 \right),
\end{equation}
and correspondingly
 \begin{equation}
\frac{\tilde{b}_\mathrm{C}}{GM} \simeq 16.72\pi A_0^2\left[\mathfrak{E}^2_\mathrm{cos}+\mathfrak{E}^2_\mathrm{sin} \right]\alpha^2\left(1-5.276 \beta_C^2 \right).
\end{equation}
In conceivable cases of GW detection where $\sqrt{A_0^2[\mathfrak{E}^2_\mathrm{cos}+\mathfrak{E}^2_\mathrm{sin}]}\sim 10^{-21}$, $\alpha\sim1/4$, and $(1-5.276\beta_C^2)\approx1$, one finds the ballpark scale $\tilde{b}_\mathrm{C}/(GM) \sim\pi\times10^{-42}$. This strong suppression arises naturally from the quadratic dependence on the GW strain amplitude -- a consequence of keeping the polarizations in the result --, which restrains us from computing a geometric scale associated with the source. Reinstating the detector-dependent factor further suppresses the computation to $b_\mathrm{C}/(GM)\sim10^{-63}$. 

On the other hand, if one had originally removed the graviton (or at this stage, the GW) polarizations after having calculated the scattering amplitude (or the classical-regime impact parameter $b_\mathrm{C}$), the remaining expression for e.g. $\tilde{b}_\mathrm{C}/(GM)$ yields values of order unity. Specifically, $\tilde{b}_\mathrm{C}/(GM)\approx1.76\pi$ under the same ballpark scalings. This calculation reflects the underlying geometric scales of the source; the value of $1.76\pi$ is much closer to the total mass horizon diameter of $\tilde{b}_\mathrm{C}/(GM)=4$, which is the separation at which coalescence ends in the mass-shell model via Eq. (\ref{rads}). It is, after all, the value to expect for the merger phase of coalescence, complementing the early-inspiral (pre-merger) $b/(GM)>14$ calculation made by WQFT.  Reinstating the detector-dependent factor gives the impact parameter scale a quintessential GW strain computation: $b_\mathrm{C}/(GM)\sim10^{-21}$. From this, we claim that $b_\mathrm{C}=(GM)\times10^{-21}$, which we may quantify with a range of ballpark orders of magnitude. 

Provided the total mass of a CCB scales with the solar mass: typically of order $\sim1M_\odot$ for binary neutron stars and at most $\sim10^2M_\odot$ for binary black holes, we use the fundamental length scale $GM_\odot=1.476$ km to quantify an expected range in the GW-mirror impact parameter: $b_\mathrm{C}\sim10^{-21}-10^{-19}$ km, or equivalently $b_\mathrm{C}\sim10^{-18}-10^{-16}$ m. These values in meters are below the nuclear scale, as $\sim10^{-15}\,\mathrm{m}=1\,\mathrm{fm}$ gauges the nucleon diameter. Indeed, the $10^{-18}$-meter length gauges the suspended mirror's recoil after making contact with a propagating GW, given a LIGO arm length of 3 km. For the ET with proposed arm lengths of 10 km, either L-shaped or as an equilateral triangle, the mirror recoil would be $\Delta L\sim10^{-17}$m, which is still within the ballpark range of $b_\mathrm{C}$.

\subsection{On leading-order Compton diagrams (Dispersive tests to GR)}\label{disps}

This work considers a tree-level graviton-scalar Compton scattering, which entails as seen in Figure \ref{fig:feyn} the external graviton and scalar lines, and a scalar propagator connected by two scalar-graviton vertices. Since we found that the $s$-channel diagram in Figure \ref{fig:feyns} produced a vanishing amplitude, we refer inherently to the $t$-channel diagram in Figure \ref{fig:feynt}. Further tests to GR include, among others, the non-zero dispersion of GWs (e.g. \cite{LIGOScientific:2017bnn}). These consider the plausible distortion of the coherent bulk motion through some dephasing in the GWs, caused by the contact with e.g. a LIGO hanging mass, let alone any object in the GW path. In the context of gravitons in a coherent-state bulk, the non-zero dispersion of GWs translates into the hypothetical release of gravitons from the coherence state, becoming ``free" particles to gauge e.g. individual graviton mass and its associating Compton wavelength, and possibly detect them \cite{Tobar:2023ksi}.

E.g. for GW170104, sourced by a quasi-circular binary black hole system, dispersive tests to GR deduced a graviton mass of $m_g\leq7.7\times10^{-23}$ eV with Compton wavelength $\lambda_g>(1.5\sim1.6)\times10^{13}$ km. These measurements, however, stem from a modified dispersion relation of the propagating GW: $E^2-|\vec{p}|^2=Ap^\alpha$, where Lorentz invariance is preserved for $A\ll1$ and $\alpha=0$. Modified theories of gravity predict various values for $\alpha$; the graviton mass and Compton wavelength were proposed under the massive-gravity parameterizations $A>0$ and $\alpha=0$ (i.e. $Ap^\alpha\rightarrow A$ effectively introduces the non-zero graviton mass $A=m_g^2$). Despite current endeavors to detect gravitons, they remain hypothetical. And as we have alluded via Eq. (\ref{gb}), detecting only one in a LIGO-like detection design is inconceivable.

Should one draw non-zero GW dispersion as a EFT Feynman diagram, the tree-level Compton interaction via Figure \ref{fig:feynt} may include, qualitatively speaking, an internal-line mechanism whereby the GW dephasing leads to graviton release. This, essentially, proposes a leading-order representation of the tree-level diagram analogous to those analyzed in QED (see Refs. \cite{Lee:2021iid, GlueX:2025hve,  Li:2025bsq}, albeit they considered next-to-leading order (NLO) and next-to-next-to-leading order (2NLO) terms in the QED Comption analysis).

Possible designs of internal-line mechanisms are provided in Figure \ref{internal}, where Figure \ref{loops} depicts a mechanism where a graviton half-loop emerges, however with no graviton release. On the other hand, Figure \ref{grad} shows a mechanism for GW dephasing via the graviton half-loop splintered into a released graviton; the half-loop therefore becomes a 3-graviton vertex. A natural consequence of introducing these mechanisms as leading-order corrections to the tree-level diagram is the increase in leading PM order: via the half-loop alone, 2PM due to four total vertices; and via the 3-graviton vertex, 2.5PM.

\begin{figure}[h!]
\centering
  \subfigure[]{\begin{tikzpicture}
\begin{feynman}
    \vertex (i);
    \vertex [above=1cm of i] (v);
    \vertex [above=1.5cm of v] (f);
    \vertex [above=1cm of f] (w);

    \diagram* {
        (i) -- [fermion, thick] (v) -- [fermion, thick] (f) -- [fermion, thick] (w),
        (v) -- [boson, half right, very thick] (f),
    };
\end{feynman}
\end{tikzpicture}\label{loops}}
	\hspace{2.5cm}
\subfigure[]{\begin{tikzpicture}
\begin{feynman}
    \vertex (i);
    \vertex [above=1cm of i] (v);
    \vertex [above=1.5cm of v] (f);
    \vertex [above right = 1.05cm of v] (o);
   \vertex [right = 1cm of o] (x);
    \vertex [above=1cm of f] (w);

    \diagram* {
        (i) -- [fermion, thick] (v) -- [fermion, thick] (f) -- [fermion, thick] (w),
        (v) -- [boson, very thick] (o),
	 (o) -- [boson, very thick] (f),
	 (o) -- [boson, very thick] (x),
    };
\end{feynman}
\end{tikzpicture}\label{grad}}
\caption{\label{internal} Prospective internal-line mechanisms to approach non-zero dispersion. (a) the $t$-channel scalar propagator has a graviton half-loop (no graviton release, sudden absorption). (b) the half-loop splinters as a released graviton, forming a 3-graviton vertex.}
\end{figure}

\section{Concluding Statements}\label{concl}

We utilized EFT to a LIGO-like graviton-scalar Compton interaction, arguably successfully. Showing that the CM energy between a GW graviton (one part in the coherent bulk with energy $E_\mathrm{g}=\omega_\mathrm{GW}$) and a suspended target with rest energy $E_M=m_M$ is of order $10^{1.5}$ PeV, this supports (and we confirm) a convergent cross section after calculating the total scattering amplitude. After one discards the remnant graviton polarization factors while keeping the kinematic information, one finds a classical regime impact parameter (after coherence state enhancement) between the incoming GW and suspended mass to be of the same quintessential order as the mirror recoil $\Delta L\sim10^{-18}$ m after making contact with the GW. One also deduces the geometric scale of the quadrupolar source e.g. at the pre-merger stage; the revised scale $\tilde{b}/(GW)\approx\pi$ approximates the coalescence separation of the CCB, provided it is modelled in the hollow mass-shell approach. 

It is our hope to motivate further useage of EFT to LIGO-like GW detections, especially for future-generation detectors such as the ET and LISA, even the Cosmic Explorer \cite{Evans:2023euw}. More specifically, leading-order corrections from the tree-level Feynman diagrams, discussed in Section \ref{disps}, are of interest to be pursued for dispersive tests of GR in the EFT perspective. Another interesting topic to eventually pursue lies in the analysis of trailing terms in the total cross section (i.e., the additional terms containing higher orders in $\omega_\mathrm{GW}/m_M$), deducing what these additional terms represent in the context of LIGO-like instrumentation, and how they influence the net worth of the total cross section and GW-mirror impact parameter beyond the first-order term analyzed in this work.

\vfill


\section*{Statement Declarations}

\subsection*{Conflict of Interest}
The author declares no conflicts of interest.

\subsection*{Data Access Statement}
As a theoretical study, this work generates no original data.

\subsection*{Ethics Statement}
No ethical issues arise, as no test subjects are involved. This paper adheres to academic integrity.

\subsection*{Funding Statement}
This work received no funding.






\end{document}